\documentclass[11pt,prl,reprint, longbibliography,superscriptaddress]{revtex4-1}

\usepackage{graphicx}
\graphicspath{{Figures/}}

\usepackage{amsmath} 
\usepackage{amsfonts}
\usepackage{bbm}
\usepackage{braket}
\usepackage{graphicx}
\usepackage{float}
\usepackage{url}
\usepackage{qcircuit}
\usepackage{amssymb}
\usepackage{qcircuit}
\usepackage{siunitx}
\usepackage{comment}
\usepackage{mwe}    % loads »blindtext« and »graphicx«

\newcommand{\yb}{$^{171}$Yb$^+$ }

\begin{document}

\title{Simulating conical intersections with trapped ions}%tentative

\author{Jacob Whitlow}
\affiliation{Duke Quantum Center, Duke University, Durham, NC 27701 USA}
\affiliation{Department of Electrical and Computer Engineering, Duke University, Durham, NC 27708, USA}
\author{Zhubing Jia}
\affiliation{Duke Quantum Center, Duke University, Durham, NC 27701 USA}
\affiliation{Department of Physics, Duke University, Durham, NC 27708, USA}
\author{Ye Wang}
\altaffiliation{Current address: School of Physical Sciences, University of Science and Technology of China, Hefei 230026, China}
\affiliation{Duke Quantum Center, Duke University, Durham, NC 27701 USA}
\affiliation{Department of Electrical and Computer Engineering, Duke University, Durham, NC 27708, USA}

\author{Chao Fang}
\affiliation{Duke Quantum Center, Duke University, Durham, NC 27701 USA}
\affiliation{Department of Electrical and Computer Engineering, Duke University, Durham, NC 27708, USA}

\author{Jungsang Kim}
\affiliation{Duke Quantum Center, Duke University, Durham, NC 27701 USA}
\affiliation{Department of Electrical and Computer Engineering, Duke University, Durham, NC 27708, USA}
\affiliation{Department of Physics, Duke University, Durham, NC 27708, USA}
\affiliation{IonQ, Inc., College Park, MD 20740, USA}
\author{Kenneth R. Brown}
\email{ken.brown@duke.edu}
\affiliation{Duke Quantum Center, Duke University, Durham, NC 27701 USA}
\affiliation{Department of Electrical and Computer Engineering, Duke University, Durham, NC 27708, USA}
\affiliation{Department of Physics, Duke University, Durham, NC 27708, USA}
\affiliation{Department of Chemistry, Duke University, Durham, NC 27708, USA}

\begin{abstract}
 Conical intersections are common in molecular physics and photochemistry, and are often invoked to explain observed reaction products. A conical intersection can occur when an excited electronic potential energy surface intersects with the ground electronic potential energy surface in the coordinate space of the nuclear positions.  Theory predicts that the conical intersection will result in a geometric phase for a wavepacket on the ground potential energy surface. Although conical intersections have been observed experimentally, the geometric phase has not been observed in a molecular system. Here we use a trapped atomic ion system to perform a quantum simulation of a conical intersection. The internal state of a trapped atomic ion serves as the electronic state and the motion of the atomic nuclei are encoded into the normal modes of motion of the ions. The simulated electronic potential is constructed by applying state-dependent forces to the ion with a near-resonant laser. We experimentally observe the geometric phase on the ground-state surface using adiabatic state preparation followed by motional state measurement. Our experiment shows the advantage of combining spin and motion degrees of freedom in a quantum simulator.
\end{abstract}

\maketitle

\newpage

%\section{Main}

% The study of molecular physics is one of the most important and long-standing applications of quantum mechanics.  It's fundamental to our understanding of chemistry and biology, and has applications towards everything from drug development and material science to precision measurement of fundamental constants in nature. Despite this, there are still many outstanding problems, and much of what we do understand relies on approximations that work beautifully in many situations, but limit our theoretical access to more complicated systems of interest.  

% One such approximation is the Born-Oppenheimer approximation, which allows chemists to take advantage of the large mass and time scale differences between nuclei and electrons in molecules and separate their quantum wavefunctions.  Oftentimes, the coordinates of the nuclei can be treated as classical variables with great accuracy, creating effective potential energy surfaces for each electronic state and significantly decreasing the Hilbert space of interest.  However, this approximation breaks down around a phenomena known as a conical intersection.  Here, the potential energy surfaces associated with two electronic states intersect, and the coupling between states due to nuclear coordinates becomes too strong to ignore the quantum effects.  The Hilbert space of such systems can become prohibitively large for classical computational methods, especially when multiple conical intersections appear near to each other.

Simulation of the quantum mechanics of molecules is an important and natural utilization of quantum simulators, with applications in calculating ground state energies and chemical reaction rates \cite{ mcardle2020quantum, kassal2008polynomial}. %\cite{feynman2018simulating, nam2020ground}.  
Classical computers have difficulty simulating the exact dynamics of even relatively simple molecules, usually resorting to an assortment of approximations to overcome the exponentially scaling Hilbert space.  The Born-Oppenheimer approximation often is used to limit the size of the Hilbert space, taking advantage of the mass differences between nuclei and electrons to separate their wavefunctions. The slow-moving nuclear positions can then be treated as parameters when calculating the energy state of the fast-moving electrons.  This allows one to visualize the movement of the nuclei on electronic state dependent adiabatic potential energy surfaces parameterized by the nuclear coordinates. This approximation breaks down when the potential energy surfaces cross at a conical intersection \cite{larson2020conical, yarkony1996diabolical}.  Near these singularities, the couplings between the nuclear and electronic coordinates become too strong to ignore. This is also where non-trivial geometric phases come into play \cite{berry1984quantal}. Such a phase depends on the direction of travel and the solid angle encompassed by the nuclear wavefunction as it makes a loop with respect to the conical intersection. This results in a phase interference not predicted by the energy dynamics of the system if different parts of the wavefunction take different paths around the intersection. Conical intersections are difficult to probe in real chemical systems due to the ultrafast and non-radiative nature of state transitions in their vicinity \cite{farag2016probing, koppel1983ultrafast, chen2019mapping}.

% They are also not unique to molecular systems, making appearances in condensed matter systems and cold-atom gasses via Dirac cones and the Rashba effect \cite{manchon2015new, juzeliunas2010generalized, armitage2018weyl}.

Quantum simulators do not run into the scaling problems that classical computers experience when performing chemical calculations \cite{kassal2008polynomial, blatt2012quantum, lloyd1996universal}.  They have already been suggested as a means of probing conical intersections and other molecular phenomena \cite{macdonell2021analog, gambetta2021exploring, wuster2011conical, wuster2018rydberg, macdonell2022predicting, omiya2022analytical, tamiya2021calculating}. Early results on calculating branching ratios have been demonstrated on superconducting systems \cite{wang2022observation}, and a similar phenomenon in condensed matter systems has been simulated with ultra-cold atom systems \cite{brown2022direct}.  Here, we explore geometric phase interference in a system based on chains of trapped ions, which are proving to be a robust and highly controllable way of simulating other quantum mechanical systems
\cite{nam2020ground, hempel2018quantum, porras2012quantum, gorman2018engineering, richerme2022quantum, monroe2021programmable, nguyen2022digital}.
Two internal states of an ion are chosen to represent a qubit, and lasers are used to coherently manipulate these states. Laser interactions that couple the qubit states to the motion of the chain are used to coherently entangle the ions with their own vibrational states. 
We utilize these vibrations to act as nuclear coordinates in a hybrid digital-analog approach to quantum simulation \cite{georgescu2014quantum, macdonell2021analog}.  We use adiabatic evolution of an ion's wavefunction to provide experimental demonstration of the creation and control of a conical intersection.  The final spatial distribution of the wavefunction is measured, exhibiting interference arising from non-trivial geometric phase.

\subsection{Model Hamiltonian and Ideal Results}

\par
\begin{figure*}[ht!]
    \begin{tabular}{cc}
        \includegraphics[width=0.90\textwidth]{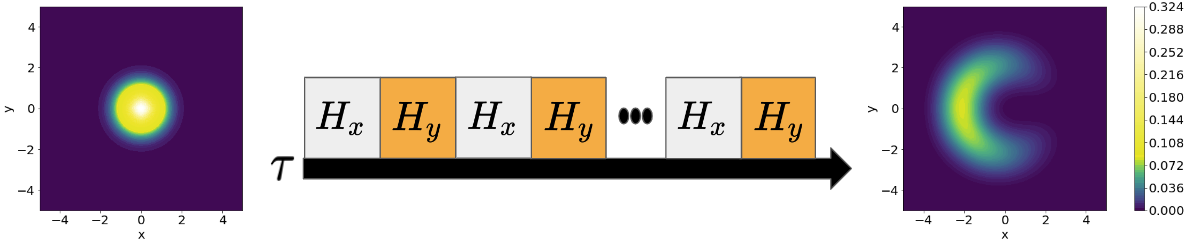}
        (a) 
    \end{tabular}
    \begin{tabular}{cc}
        \includegraphics[width=0.25\textwidth]{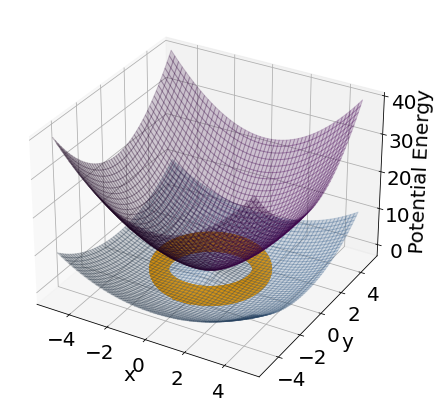} 
        (b) 
        \includegraphics[width=0.2\textwidth]{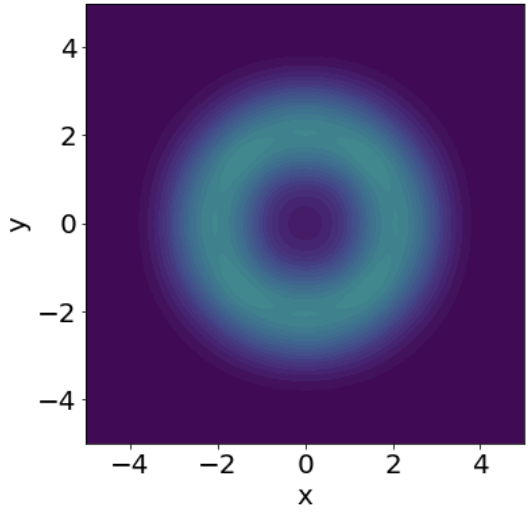} 
        (c) 
        \includegraphics[width=0.2\textwidth]{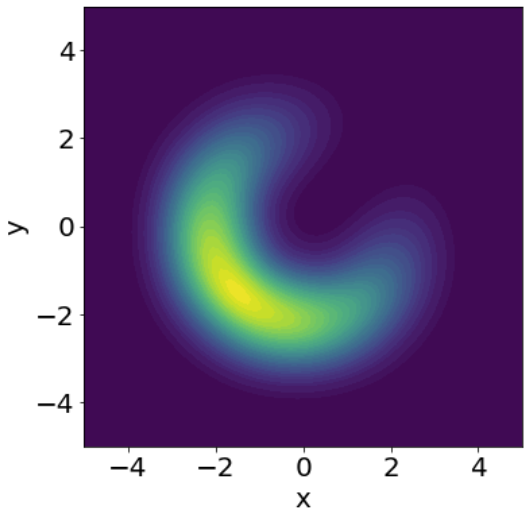} 
        (d) 
    \end{tabular}
    \caption{\textbf{a}, The Trotterized adiabatic evolution scheme which switches between $x$-mode ($H_x$) coupling and $y$-mode ($H_y$) coupling, labeled. The plots show the probability distributions of the wavefunction in the $x$ and $y$ modes at the beginning and the end of the evolution, based on classical simulations.  Note that all classical simulations were done with the open source package QuTiP \cite{johansson2012qutip}. They assume no error from Trotterization or non-adiabatic behavior. \textbf{b}, The potential energy surface produced by the ideal Hamiltonian in Eq. \ref{equation:semi_classic} when setting $m$ and $\nu$ to $1$, and $\Omega$ to $3$, revealing a conical intersection.  The orange torus surrounding the intersection represents the spatial distribution of the ground state subspace of the Hamiltonian. \textbf{c}, The probability distribution of the final state if there were no geometric interference, allowing the wavefunction to meet with itself on the other side of the intersection. \textbf{d}, The effect of a $\frac{\Delta}{2}\sigma_z$ term on the final state is a rotation.  The first order effect is to rotate the state by $\Delta \tau$, where $\tau$ is the time of the evolution.}
    \label{fig:Energy_Surface}
\end{figure*}

Initially, we consider an ideal Hamiltonian of the form:
\begin{equation}
    \hat{H}_{ideal} = \frac{\nu}{2}\big(p_x^2 + p_y^2 + x^2 + y^2\big) + \frac{\Omega}{\sqrt{2}} \big(\hat{\sigma}_x x + \hat{\sigma}_y y\big).
    \label{equation:semi_classic}
\end{equation}
This describes a spin-$\frac{1}{2}$ particle in a two-dimensional harmonic oscillator with vibrational frequency $\nu$, where the spin of the particle is coupled to its own position via the Pauli operators $\sigma_x$ and $\sigma_y$ with a strength $\Omega$.   The dimensionless positions, $x$ and $y$, and momenta, $p_x$ and $p_y$, are normalized by a factor of $\frac{1}{\sqrt{m \nu}}$ and $\sqrt{m \nu}$ respectively, where we have set $\hbar=1$ throughout the manuscript. The positions and momenta are first considered as parameters for the spin Hamiltonian, as in the Born-Oppenheimer approximation. 
% Note that these are dimensionless positions and momentums, normalized by a factor of $\frac{1}{\sqrt{m \nu}}$ and $\sqrt{m \nu}$ respectively. 
The energies of the spin eigenstates, which correspond to the electronic eigenstates, are coupled in opposite ways to the positions in the harmonic oscillators.  The eigenenergies for the higher (+) and lower ($-$) states in this system are $E_{\pm}(x, y) = \frac{\nu}{2}\big(p_x^2 + p_y^2) + V_{\pm}(x, y)$, where $V_{\pm}(x, y) = \frac{\nu}{2}(x^2 + y^2\big) \pm \frac{1}{\sqrt{2}} \sqrt{\Omega^2 x^2 + \Omega^2 y^2}$ is the spin-dependent potential energy of the system. A plot of $V_{\pm}(x, y)$ is shown in Fig. \ref{fig:Energy_Surface}b, where a conical intersection can clearly be seen.  

\par
An important feature of this semi-classical Hamiltonian is the possibility of a geometric phase that depends only on the movement through space. If the evolution of the system is adiabatic enough (see Methods section), a phase that is entirely separate from the energy-dependent one that accumulates over time will appear \cite{berry1984quantal, berry1990anticipations}.  If $x$ and $y$ were time dependent parameters in this system ($x(t)$ and $y(t)$), and the system were to start in one of the eigenstates of the system with the energies described above, we would see the following adiabatic time evolution:
\begin{equation}
    \begin{aligned}
        |\psi(t) \rangle & = T \exp \bigg(-i\int^t dt' \hat{H}(t') \bigg) |\psi_{n, 0} \rangle \\ 
        & = e^{-i\int^t dt' E_n(R(t'))} e^{i \gamma_n(t)} |\psi_n(R(t)) \rangle,
    \end{aligned}
\end{equation}
where $T$ is the time ordering operator, $R(t) = (x(t), y(t))$,  $E_n(R(t))$ is the position dependent energy of the $n$th eigenstate $|\psi_n(R(t)) \rangle$, and $\gamma_n(t)$ is the geometric phase associated with that state, described by the following equation:
\begin{align}
    % \begin{aligned}
        \gamma_n & = i\int^t dt' \langle \psi_n(R(t'))|\nabla_R \psi_n(R(t'))\rangle \cdot \dot{R}(t').\\
        & = i\int^{R_f}_{R_0} dR' \cdot \langle \psi_n(R')|\nabla_{R'} \psi_n(R')\rangle.
    % \end{aligned}
\end{align}
Given that the spin eigenstates as a function of position can be written as $\ket{\pm (x, y)} = \frac{1}{\sqrt{2}}(\ket{0} \pm \frac{x + iy}{\sqrt{x^2 + y^2}}\ket{1})$, this amounts to the following integral:

\begin{equation} \label{equation:xy_form}
    \gamma_{\pm} = -\frac{1}{2}\int^{x_f, y_f}_{x_0, y_0}  \frac{y dx - x dy}{x^2 + y^2}.
\end{equation}
Berry originally pointed out that if the path taken through this position space were to perform a closed loop $C$ around a degeneracy point, the phase would be equal to one half of the solid angle, $\tilde{\Omega}$, subtended by the loop around that point: $\gamma(C) = \frac{1}{2}\tilde{\Omega}(C)$. Wavepackets that travel in opposite directions around the degeneracy point acquire opposing geometric phases which interfere destructively, an effect we experimentally verify in this work.

Another parameter that can be added to our system is the energy difference between the $\ket{\uparrow}$ and $\ket{\downarrow}$ spin states, $\Delta$.  This would add $\frac{\Delta}{2}\sigma_z$ to our Hamiltonian, which creates an avoided crossing in the system where the conical intersection would be. If $\Delta \ll \Omega$, this type of system is still highly non-adiabatic in behavior and also allows for geometric phase interference based on the solid angle argument, where the solid angle is now $2 \pi \big(1 - \frac{\Delta}{\sqrt{\Omega^2 + \Delta^2 }}\big)$ \cite{berry1984quantal, berry1990anticipations}. The topological nature of this system is less obvious however, because we are missing the singularity associated with the intersection.  However, by moving into the rotating spin frame, we bring back the intersection with the new Hamiltonian:

\begin{equation}
    \hat{H} = \hat{H}_{H.O} + \frac{\Omega}{\sqrt{2}}(e^{-i\Delta t}(x - iy)\sigma_- + h.c.)
    \label{equation:semi_classic_rotating}
\end{equation}
Here, $\hat{H}_{H.O.}$ is the classical 2D harmonic oscillator, $\sigma_{\pm} = \frac{1}{2}(\sigma_x \mp i \sigma_y)$, and $h.c.$ is the hermitian conjugate. The potential energy surface is the same as in Eq. \ref{equation:semi_classic}, but the axes are rotating. When the energy difference is small compared to the strength of the coupling, this has the effect of adiabatically moving the wavefunction around the intersection. The geometric phase is still there, as is the topological nature of the system. 
% In the original frame, as $\Delta$ is increased, the energy surfaces move further away, the system becomes more adiabatic and the solid angle encompassed by the closed loop vanishes.  In this rotating frame, the intersection is always there, but without synchronizing the experiment to this frame the faster speed of rotation quickly washes out any geometric phase interference on average.  

\par
A more accurate model requires coupling between the nuclear and electronic quantum states because of the proximity to the intersection and breakdown of the Born-Oppenheimer approximation. Assuming vanishing $\Delta$, this leads to the following fully quantum Hamiltonian:
\begin{equation}
    \hat{H}_{ideal} =  \nu \hat{n}_x +  \nu \hat{n}_y + \frac{ \Omega}{2} \hat{\sigma}_x (\hat{a}_x + \hat{a}_x^{\dagger}) + \frac{ \Omega}{2} \hat{\sigma}_y (\hat{a}_y + \hat{a}_y^{\dagger}).
    \label{equation:quantum_JT}
\end{equation}
Here, we have replaced $\xi$ and $p_{\xi}$ by ladder operators $\frac{1}{\sqrt{2}} (\hat{a}_{\xi} + \hat{a}_{\xi}^{\dagger})$ and $\frac{-i}{\sqrt{2}} (\hat{a}_{\xi} - \hat{a}_{\xi}^{\dagger})$ respectively, where ${\xi} \in \{x, y\}$. This Hamiltonian is associated with the Jahn-Teller effect, or Rashba coupling if position is replaced by momentum in the spin-motion coupling, as is common in condensed matter physics \cite{ longuet1958studies, manchon2015new, Lin_2011}.
%juzeliunas2010generalized,rashba2015symmetry,jahn1937stability,
The subspace of degenerate ground states for this Hamiltonian forms a ring around the center of the harmonic oscillator as one would expect from the energy surfaces plotted in Fig. \ref{fig:Energy_Surface}b.  Importantly, the arguments for geometric phase still apply in this fully quantized picture, just with a larger Hilbert space.

% The natural choice for such a path would be to first displace the wavefunction to the ground state ring's radius utilizing just the x-mode coupling, maintaining a Gaussian distribution, then open the y-mode coupling adiabatically to allow a semi-circular path around the intersection. This can be described temporally by:
% \begin{widetext}
% \begin{equation}
%     \hat{H}(t) = \frac{\Delta}{2}\hat{\sigma}_z + \nu \hat{n}_x +  \nu \hat{n}_y + 
%     \left\{
%         \begin{array}{lr}
%             \frac{ \Omega t}{2\tau} \hat{\sigma}_x (\hat{a}_x + \hat{a}_x^{\dagger}), & \text{if } 0 \leq t \leq \tau\\
%             \frac{ \Omega}{2} \hat{\sigma}_x (\hat{a}_x + \hat{a}_x^{\dagger}) + \frac{ \Omega (t - \tau)}{2\tau} \hat{\sigma}_y (\hat{a}_y + \hat{a}_y^{\dagger}), & \text{if } \tau < t \leq 2\tau
%         \end{array}
%     \right\}
% \end{equation}
% \end{widetext}
% This is demonstrated in the simulation results shown in Fig.\ref{fig:Energy_Surface}b, but was not the path we chose. 

The essence of our experiment is as follows. Start with a qubit in the eigenstae of $\sigma_x$, $\ket{+}$, and in the ground state of a 2D harmonic oscillator.  Then adiabatically turn on coupling to the motional modes of the harmonic oscillator in such a way that the path taken by the wavefunction splits up and meets on the other side. Ideally, we turn on both the motional couplings at once, as described by:

\begin{equation}
    \hat{H}(t) = \nu \hat{n}_x +  \nu \hat{n}_y + 
            \frac{ \Omega t}{2\tau} \hat{\sigma}_x (\hat{a}_x + \hat{a}_x^{\dagger}) + \frac{ \Omega t}{2\tau} \hat{\sigma}_y (\hat{a}_y + \hat{a}_y^{\dagger})
\end{equation}
Here $\tau$ is the time of the evolution. 
Due to the symmetry breaking of starting in the $\ket{+}$ state, there is an initial push to the negative x direction, followed by a path that encircles the intersection (see Supplementary Material). Experimentally, we performed the evolution in Trotterized fashion, breaking the coupling to the $x$-mode and $y$-mode into separate steps in time. The evolution scheme of this process is shown in Fig. \ref{fig:Energy_Surface}a.  This does add noise depending on the number and length of steps (see Methods). The final ideal state of the trajectory is a crescent shape due to interference on the other side of the conical intersection. Importantly, this interference must be a result of geometric phase because the system will remain in the ground state throughout the adiabatic evolution, leading to a global energy-dependent phase. This is evident in Fig. \ref{fig:Energy_Surface}c, where a Hamiltonian with only the bottom portion of the potential energy is used simulate the evolution, resulting in a ring. In Fig. \ref{fig:Energy_Surface}d, we show the results if $\frac{\Delta}{2}\sigma_z$ were included, where $\Delta \ll \Omega, \nu$.  The wavefunction is rotated slightly as predicted by moving into the rotating frame in Eq. \ref{equation:semi_classic_rotating}.

\begin{figure*}[ht!]
    \begin{tabular}{cc}
         \includegraphics[width=0.45\textwidth]{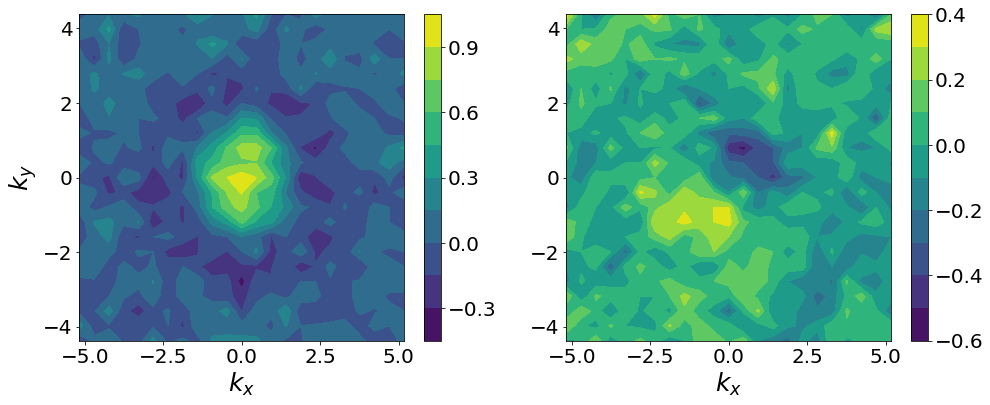} 
         (a) 
         \includegraphics[width=0.45\textwidth]{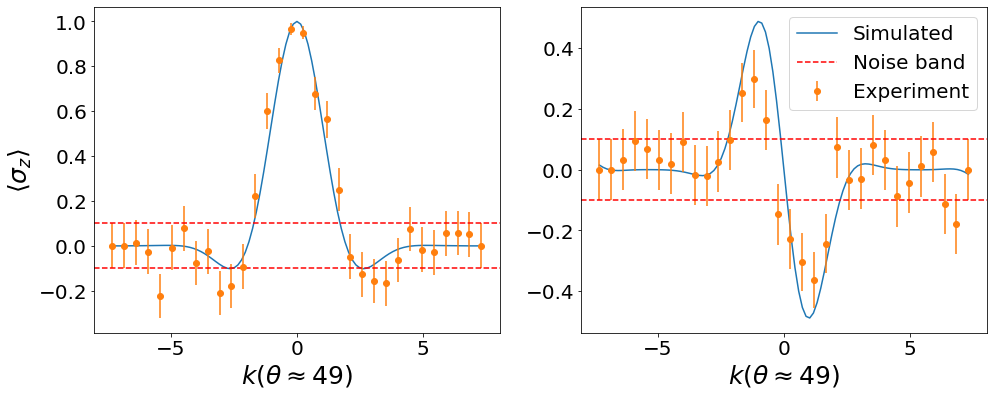} 
         (b) 
    \end{tabular}
    \caption{\textbf{a}, Result of Fourier push on second ion, after adiabatic evolution on first ion, by starting in a $\sigma_z$ eigenstate (left) and $\sigma_y$ eigenstate (right). The plots show measured values of $\braket{\sigma_z}$ based on 100 experiments for each data point. Axes are the k-vectors associated with the push, calculated based on experimental laser interaction strengths. \textbf{b}, Cutout of experimental data along a line 49 degrees to the horizontal, with $\sigma_z$ eigenstate (left) and $\sigma_y$ eigenstate (right), compared to simulated data. Shot noise dominates our measurement uncertainty and the uncertainty for an expected value of zero is indicated by the red dashed lines.   Error bars are calculated based on binomial distribution with 100 shots per data point. %Rotation of the data was done using Scipy rotate \cite{2020SciPy-NMeth}.
    }
    \label{fig:fourier_push}
\end{figure*}

\begin{figure*}[ht!]
    \begin{tabular}{cc}
         \includegraphics[scale=0.33]{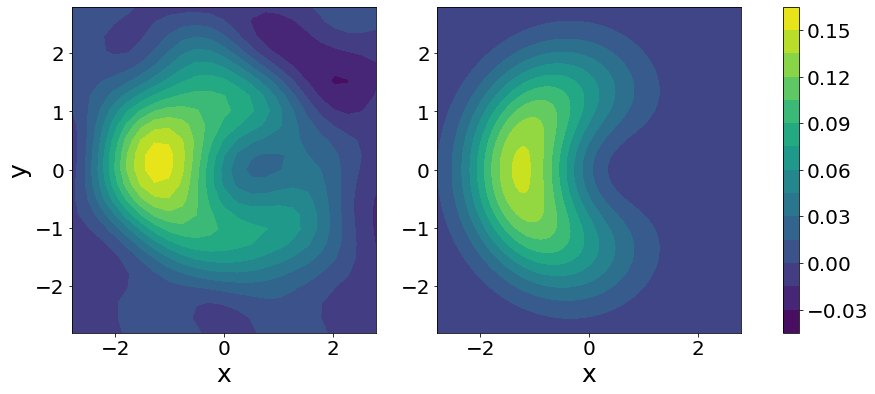} 
         (a) 
         \includegraphics[scale=0.42]{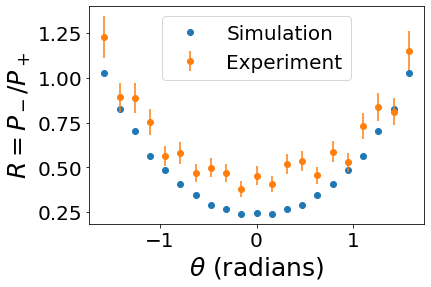} 
         (b) 
    \end{tabular}
    \caption{\textbf{a}, Experimental results (left) and classical simulation results (right) after adiabatic evolution of ion chain's wavefunction around a conical intersection in normal coordinate space, where $\nu = 2 \pi \times 3$ kHz and $\Omega = 2 \pi \times 6$ kHz. The experimental data was obtained via inverse Fourier transform of the raw experimental data rotated by 49 degrees. In theory this should be a pure probability distribution, but experimental error and measurement statistics allow for negative values.  The simulation results produce a pure probability distribution \textbf{b}, The integral of all of the distribution on one half of the plane is divided by the integral over the other half.  The line that defines the halfway point is rotated by the angle $\theta$. The plot shows a comparison of the distribution ratios vs. angle of the dividing line for classical simulation and experiment. Error bars are calculated based on propagation of error through the inverse Fourier transform, then a Taylor expansion when taking the ratios.}
    \label{fig:experimental_results}
\end{figure*}

\subsection{Experimental Results}

We performed the experiment on a room temperature trapped ion system, calibrated to trap five \yb ions \cite{wang2020high, jia2022determination}.  The atomic states and motional modes are manipulated with laser-based interactions to mimic the desired Hamiltonian (see Methods). The normal coordinate distribution was measured using a ``Fourier push'' method \cite{gerritsma2010quantum}.  By using a second ion that couples into the same modes we used for the simulation, we can perform two separate spin-dependent pushes to create the unitary evolution $U(k_x, k_y) = \exp\big(-\frac{i}{2}\sigma_x(k_x \hat{x} + k_y \hat{y})\big)$. We then extract the Fourier transform of the spatial distribution by measuring the population in the excited state of the ion and converting to $\braket{\sigma_z}$ (see Methods).  The even (odd) information is obtained by starting the second ion in a $\sigma_z$ ($\sigma_y$) eigenstate.  The results of these measurements are shown in Fig. \ref{fig:fourier_push}a. The results have been rotated by about 49 degrees due to a detuning from the resonant transition, most likely caused by AC Stark shifts, adding an effective energy difference in the spin states. In Fig. \ref{fig:fourier_push}b, we see a cutout of this data along the 49 degree line, where the most prominent features are present, in comparison to classical simulation. After $k \approx 2.8$, the data evens out at about 0, which corresponds to the point of highest shot noise when making measurements. We cutoff at this point when taking the inverse Fourier transform to avoid high frequency features due to the shot noise.

The result of taking the inverse Fourier transform of the measured data is shown in Fig. \ref{fig:experimental_results}a, where a clear crescent shape emerges after the adiabatic evolution. We rotated the axes by the necessary 49 degrees to match the high density region with the ideal evolution. This is effectively a post-measurement AC Stark shift compensation. In this experiment, we chose $\nu = 2\pi \times 3$ kHz and $\Omega = 2\pi \times 6$ kHz, and we determine our Stark shift to be around $\Delta = 2 \pi \times 0.7$ kHz based on simulations. These numbers were subject to drift of about $10 \%$ over the course the day, resulting in some experimental uncertainty.  We chose an experimental time of 330 $\mu$s, which does not satisfy the criteria for adiabaticity for a linear ramp based only on energy level spacing.  However, due to symmetries in the system, some transitions are forbidden and the effective splitting between states is higher (see Supplementary Material). Still, because our experiment only approximately met the adiabatic condition, there were slight oscillations due to non-adiabatic transitions. The time of 330 $\mu$s is also perfectly timed with the most outward point of the initial oscillation (i.e. $\frac{2 \pi}{\nu}$), making the crescent shaped wavefunction the most visible. Note that no post-processing of the data was done other than normalization of the distribution. The results should be a probability distribution in theory, but negative quantities appear in the far corners due to measurement statistics and experimental error from drift over the course of the experiment.  

Our measured distribution meets all of the qualitative features we are looking for, specifically region of maximum and minimum density on a ring-like shape.  We also compare it to the ideal results using the ratio of distribution on one half of the plot to that of the other, adjusting the angle $\theta$ of the line that defines the halfway point.  This is described mathematically by:

\begin{equation}
    R(\theta) = \frac{\int_{+} dx dy P_{\theta}(x, y)}{\int_{-} dx dy P_{\theta}(x, y)}.
\end{equation}
Here, $P_{\theta}(x, y)$ is the spatial distribution of the wave packet and $\theta = 0$  corresponds to a vertical line. The $+$ and $-$ sign indicate whether the integral was taken on the positive or negative side of the dividing line. Ideally, almost all of the distribution is on one side of the plot for $\theta = 0$, and this ratio goes to one as $\theta$ approaches $\frac{\pi}{2}$.  Experimentally, we see this general pattern but shifted up due to experimental noise. This noise can come from many things, including non-adiabatic dynamics, off-resonant coupling into other modes, motional phase mismatch, frequency drifts in our motional modes, our system not starting in the harmonic oscillator ground state, our system heating over time, and system preparation and measurement (SPAM) error.  The qualitative agreement shows the robustness of this effect to noise.

\subsection{Discussion and Future Directions}

In this work, we provide experimental evidence for geometric phase interference in simulated conical intersections in trapped ions.  We utilize spin-dependent laser pushes to create an adiabatic potential energy surface that mimics that of a small molecule.  We adiabatically evolve the ion chain under the Hamilontian, then measure interference in the spatial distribution due to the accumulated geometric phase. This should push forward work on quantum simulation of chemical systems with trapped ions to include systems with strong nuclear-electronic interactions. 

The next step is to utilize conical intersections to work towards quantum advantage for a practical problem in quantum chemistry. One can imagine a system that utilizes multiple ions to create more complicated potential energy surfaces with more than one crossing between more than two levels. There are also many more motional modes available in large ions chains, meaning we can create high-dimensional potential energy surfaces with multiple crossings, quickly escalating the necessary Hilbert space while maintaining simulability on trapped ion systems.  Work has also been done to utilize the second-order sidebands in trapped ion systems for entangling gates \cite{katz2022programmable, katz2022programmable2}. Such operations increase the complexity of the potential energy surfaces available by adding quadratic terms.  We can also utilize the bosonic modes as bath modes in an open quantum system \cite{macdonell2021analog, lemmer2018trapped}, allowing us to take advantage of decoherence as a tool which mimics noise in the environment.

There are also tools from digital quantum simulation that can be utilized in analog simulations.  Frequency and amplitude modulated pulses can be used to couple into specific modes while decoupling from others, overcoming the problem of mode-crowding and off-resonant coupling \cite{roos2008ion, leung2018robust}.  Fermionic degrees of freedom can be brought in via the Jordan-Wigner or Bravyi-Kitaev transformations, well-established techniques that map Pauli operators to Fermionic ones \cite{batista2001generalized,  seeley2012bravyi}. Trotterization can be used to string these operations together with a controllable amount of error. We can combine these with the non-adiabatic bosonic models simulated in this paper to create molecular Hamiltonians of arbitrary size and complexity, if the number of ions and the coherence times allow for it.  All of this can scale in a polynomial way with the number of degrees of freedom we wish to simulate \cite{macdonell2021analog, kassal2008polynomial}, opening up the path to practical quantum advantage for scientific computation. 
%Composite pulses can be used to improve the quality of digital quantum gates, which can be used in conjunction with analog operations \cite{brown2004arbitrarily, merrill2014progress}. 

We have recently become aware of work done by the University of Sydney on a similar experiment with a single trapped ion quantum simulator \cite{valahu2022direct}.  Their experiment explores the effects of geometric phase on the dynamics of a wavefunction as it travels around an engineered conical intersection.  This compliments our own experiment as it demonstrates how the time-dependent behavior of a system not in an eigenstate of the Hamiltonian can be affected by geometric phase in non-trivial ways. 

\section{Methods}

\subsection{Experimental Setup and Trapped Ion Hamiltonian} 

The system utilizes the hyperfine states of the ion as its qubits, i.e. $\ket{F = 0, m_F = 0} = \ket{0}$ and $\ket{F = 1, m_F = 0} = \ket{1}$ of the $^2S_{1/2}$ manifold, split by ($2\pi$)12.6428 GHz \cite{olmschenk2007manipulation}.  The ions sit in pseudo-harmonic potential on a micro-fabricated linear Paul trap that can be modelled as a quantum harmonic oscillator \cite{revelle2020phoenix}. We use 370 nm light resonant with the $^2S_{1/2} \rightarrow ^2P_{1/2}$ transition to perform Doppler cooling, detection, and initialization.

\begin{figure}[ht!]
     \includegraphics[width=0.25\textwidth]{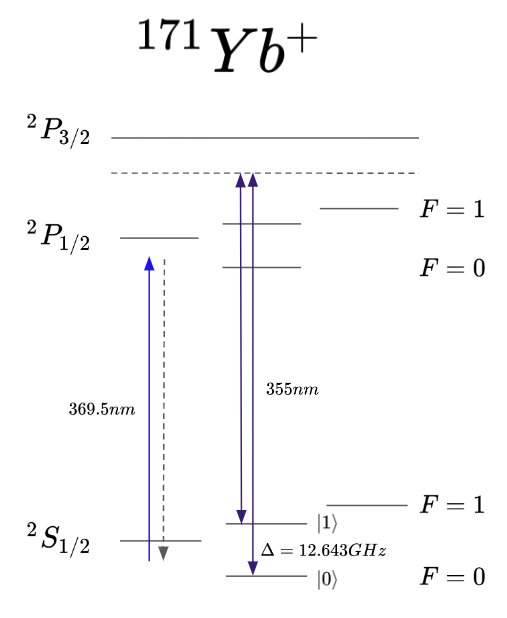} 
     (a) 
     \includegraphics[width=0.24\textwidth]{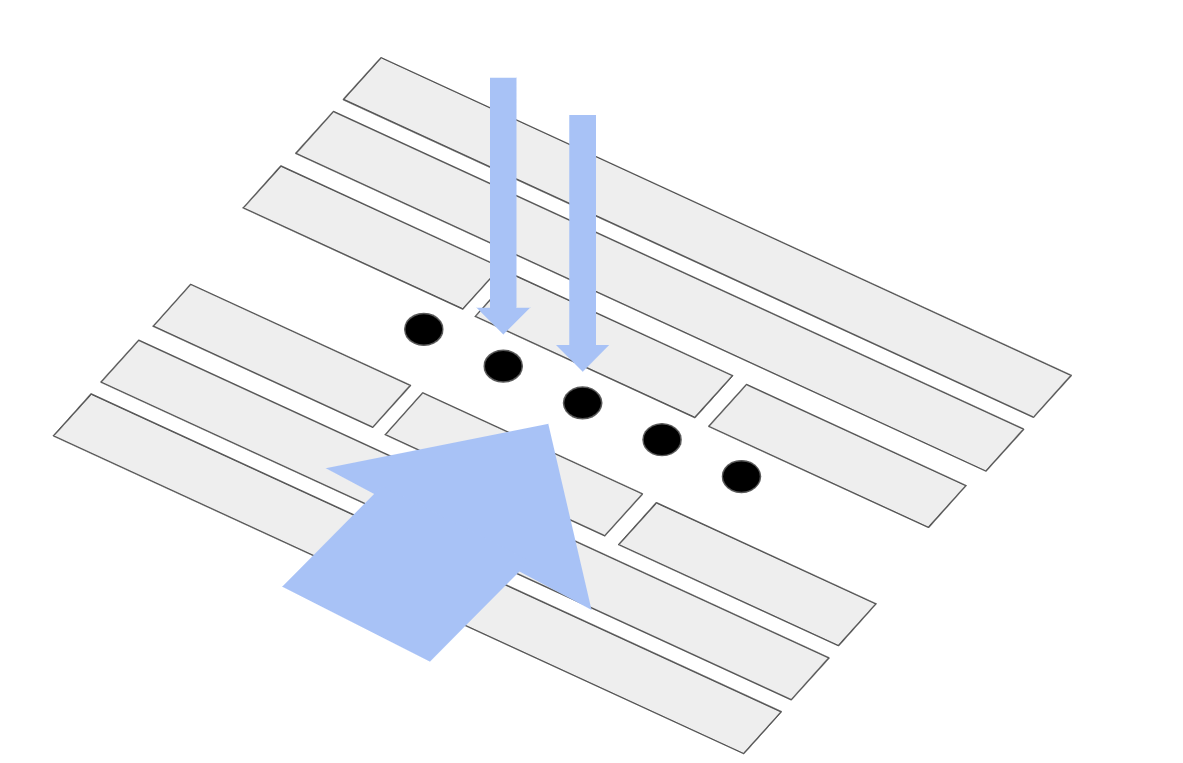}
     (b)
     \includegraphics[width=0.45\textwidth]{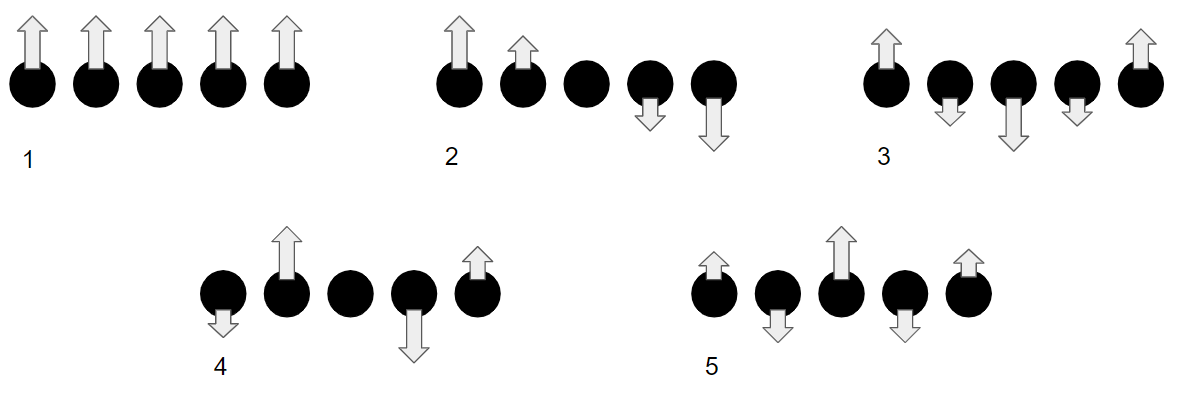}
     (c) 
    \caption{\textbf{a}, Important energy levels for cooling, detection, and Raman transitions in \yb. \textbf{b}, Layout of surface trap, five ions, global Raman beam, and individual addressing Raman beams in experimental setup. \textbf{c}, Relative coupling into the motional modes by each ion, in order of descending resonant frequency \cite{jia2022determination, debnath2016programmable}.  The ions we used were the center one for simulation, and one of its nearest neighbors for measurement.  The motional modes we used were the third and the fifth (also known as the zig-zag mode).}
    \label{fig:trap_stuff}
\end{figure}

Transitions between qubit state are made via Raman transition with two tones of a pulsed 355 nm laser \cite{debnath2016programmable, hayes2010entanglement}.
% One tone is large and elliptical, addressing all ions at once.  The other tone is delivered via tightly focused beams which individually address ions.  The pulsing creates a frequency comb separated by the repetition rate of the laser (about 118 MHz), allowing us to make up most of the frequency difference between the two states by utilizing the 106th harmonic of the comb. The rest of the difference is made up via RF signals applied to AOMs.  
Such transitions can be coupled to the motion of the ion by detuning one of the laser tones from the resonant transition frequency by exactly the frequency of the harmonic motion of the ion $\nu$. The detuning can be negative or positive, respectively referred to as red or blue sideband transitions. The effective Hamiltonians in the interaction frame for each tone can be written as:
% (see Methods):

\begin{equation}
    \begin{aligned}
        \hat{H}_r = \nu \hat{n} + \frac{\Omega \eta}{2} (\sigma_+ a e^{-i \phi} + \sigma_- a^{\dagger} e^{i \phi}), \\
        \hat{H}_b = \nu \hat{n} + \frac{\Omega \eta}{2} (\sigma_- a e^{-i \phi} + \sigma_+ a^{\dagger} e^{i \phi}).
    \end{aligned}
\end{equation}
The parameters $\nu$, $\Omega$, $\eta$, and $\phi$ are the detuning from the motional transition, the Rabi frequency, the Lamb-Dicke parameter, and the phase of the laser, respectively. One can completely recreate Eq. \ref{equation:quantum_JT} in the interaction picture of the system by utilizing many tones on the same AOMs that couple into two modes.

Because we trap five ions in this experiment, we are no longer dealing with a distribution over real space but instead over normal-mode coordinates.  This has no effect on the distribution we measure, but it allows us to use noise-insensitive modes for our experiment.  A single ion only has a center of mass mode to couple to, known to be the most susceptible to uniform electric field noise \cite{ wineland1998experimental2}.  In our five ion chain, we use the third highest frequency mode and the lowest frequency mode (also called the zig-zag mode), shown in Fig.\ref{fig:trap_stuff}c. 

% Due to calibration for two tones at a time, we performed the evolution in Trotterized fashion, breaking the coupling to the effective "x-mode" and "y-mode" into separate pulses in time \cite{trotter1959product}.  This does add noise depending on the number and length of steps, the analysis of which can be found in the Methods section. 

\subsection{Trotterized Evolution and Adiabatic Criteria}

Our system is calibrated to apply up two tones to our AOMs, allowing us to couple to one mode at a time. We  break the evolution of the system under the Hamiltonian into steps, commonly called Trotterizaion, given by the following equation:

\begin{equation}
    e^{-i t (\hat{H}_x + \hat{H}_y)} = e^{-i t \hat{H}_x}e^{-i t \hat{H}_y} + O(t^2)
 \end{equation}
By making $t$ small enough for each evolution, we can remove most error.  In our experiment, we broke up the Hamiltonian into the following way:

\begin{equation}
    \begin{aligned}
        \hat{H}_x = \nu_x \hat{n}_x + \frac{\Omega_x}{2}\sigma_x (\hat{a}_x + \hat{a}^{\dagger}_x), \\
        \hat{H}_y = \nu_y \hat{n}_y + \frac{\Omega_y}{2}\sigma_y (\hat{a}_y + \hat{a}^{\dagger}_y).
    \end{aligned}
\end{equation}
The harmonic oscillator term $\nu\hat{n}$ can be simulated with a calibrated laser detuning or by adding phase proportional to the time of the interaction to the laser pulses at each step. We chose the latter because we found it easier to control at the cost of a temporal digitization of the oscillator term.  This effectively breaks the Trotterization into four steps, with the oscillator term and the spin-dependent push treated separately. We performed each evolution sequentially 16 times, with adiabatically increasing laser strengths. This was found to incur a negligible error, based on classical simulations.

In order to minimize the error from decoherence, we also wished to shorten our experiment as much as possible while maintaining the following adiabatic condition for all times $t$ \cite{larson2020conical}:

\begin{equation}
    \bigg|\frac{\bra{\psi(t)_n}\frac{d}{dt}\hat{H}(t) \ket{\psi(t)_m}}{\Delta_{nm}^2(t)}\bigg| \ll 1,
    \label{equation:adiabatic_condition}
\end{equation}
where $\ket{\psi_n(t)}$ and $\ket{\psi_m(t)}$ are states in the nth and mth eigensubspace at time $t$, respectively, and $\Delta_{nm}$ is the difference in energy between these two subspaces.  Often overlooked in this equation is that $\frac{d}{dt}H(t)$ needs to couple the two subspaces.  For our chosen experimental time of 330 $\mu s$, the difference in frequency does not actually qualify for a simple energy argument for adiabaticity.  Due to symmetries however, the lowest subspaces are not coupled together, and this condition is met to good approximation. In fact, simulations show that the value on the left side of inequality (\ref{equation:adiabatic_condition}) never goes above 0.5 for the first 8 excited states, and is consistently below 0.2. 

% \begin{figure}[ht!]
%     \includegraphics[scale=0.5]{Figures/trotter_simulated.png}
%     \caption{The simulated results of a Trotterized evolution with an overall evolution time of 330 $\mu s$, broken into 16 steps, when turning on both x and y interactions at the same time.}
%     \label{fig:trotter_simulated}
% \end{figure}

% The simulated final state of the Trotterized adiabatic evolution is shown in Fig.\ref{fig:trotter_simulated}.  The fidelity with the desired final state based on the naive evolution and an evolution time of 25 $ms$ is about 83\%, but most importantly the phase interference is still clearly visible.

\subsection{Measurement Procedure}

\par
In order to confirm the success of our experiment, we need to measure the spatial distribution of the ion's wavefunction.  To do this, we expand the 1D ``Fourier push'' applied in \cite{gerritsma2010quantum} to 2D, taking advantage of the fact that we have multiple ions in our setup, many of which are unused during the experiment and therefore their internal states are not entangled with the motion.  We choose one that couples into the same modes as the experimental ion and perform a state dependent push on it defined by:

\begin{equation}
    U(t_1, t_2) = e^{-\frac{i\Omega \eta}{2}\sigma_x (t_1 \hat{a}_x + t_2 \hat{a}_y + t_1 \hat{a}_x^{\dagger} + t_2 \hat{a}_y^{\dagger})}.
\end{equation}
Note that the ion is pushed in the two different spatial directions for different amounts of time, $t_1$ and $t_2$.  This is easy enough to accomplish because pushes in different directions commute with each other, so they can be performed one after another.  This can be recast as:

\begin{equation}
    U(k_x, k_y) = e^{-\frac{i}{2}(k_x x + k_y y)}.
\end{equation}
Some simple algebra tells us that measuring the state of this ion in the $\sigma_z$ basis provides the following averages:
\begin{equation}
    \begin{aligned}
        \langle U(k_x, k_y)^{\dagger} \sigma_z U(k_x, k_y) \rangle = & \langle \sigma_z \cos(k_x x + k_y y) \rangle + \\ & \langle \sigma_y \sin(k_x x + k_y y) \rangle.
    \end{aligned}
\end{equation}
Therefore, by performing the same experiment twice but preparing the extra ion in the positive eigenstate of the $\sigma_z$ operator first and the $\sigma_y$ operator second, we can construct the Fourier transform of the spatial distribution.

\subsection{Acknowledgments} We thank Christophe Valahu, Vanessa Olaya-Agudelo, Ting Rei Tan, Ivan Kassal, Michael Biercuk, and Ely Novakoski for insightful discussions.
This work was supported by the Office of the Director of National Intelligence, Intelligence Advanced Research Projects Activity through ARO Contract W911NF-16-1-0082, the National Science Foundation STAQ Project Phy-181891, and the U.S. Department of Energy, Office of Advanced Scientific Computing Research QSCOUT program,  DOE Basic Energy Sciences Award No. DE-0019449, ARO MURI Grant No. W911NF-18-1-0218, and the NSF Quantum Leap Challenge  Institute for Robust Quantum Simulation Grant No. OMA-2120757.
% As mentioned in \cite{macdonell2021analog}, pyrazine is a molecule exhibiting conical intersections that could allow for a simulation on just a single ion. Other potential near-term molecules to simulate could be carbon monohydrides, which play important roles in combustion and interstellar chemistry and exhibit non-adiabatic behavior with avoided crossing and conical intersections \cite{laws2022velocity},  and perhaps benzene cations, which play important roles in organic chemstry \cite{galbraith2017few, tachikawa2018jahn, baldea2007jahn}.
 
\section{References}

\bibliographystyle{naturemag} % Tell bibtex which bibliography style to use
\bibliography{refs.bib}

\begin{thebibliography}{10}
\expandafter\ifx\csname url\endcsname\relax
  \def\url#1{\texttt{#1}}\fi
\expandafter\ifx\csname urlprefix\endcsname\relax\def\urlprefix{URL }\fi
\providecommand{\bibinfo}[2]{#2}
\providecommand{\eprint}[2][]{\url{#2}}

\bibitem{mcardle2020quantum}
\bibinfo{author}{McArdle, S.}, \bibinfo{author}{Endo, S.},
  \bibinfo{author}{Aspuru-Guzik, A.}, \bibinfo{author}{Benjamin, S.~C.} \&
  \bibinfo{author}{Yuan, X.}
\newblock \bibinfo{title}{Quantum computational chemistry}.
\newblock \emph{\bibinfo{journal}{Rev. Mod. Phys.}}
  \textbf{\bibinfo{volume}{92}}, \bibinfo{pages}{015003}
  (\bibinfo{year}{2020}).

\bibitem{kassal2008polynomial}
\bibinfo{author}{Kassal, I.}, \bibinfo{author}{Jordan, S.~P.},
  \bibinfo{author}{Love, P.~J.}, \bibinfo{author}{Mohseni, M.} \&
  \bibinfo{author}{Aspuru-Guzik, A.}
\newblock \bibinfo{title}{Polynomial-time quantum algorithm for the simulation
  of chemical dynamics}.
\newblock \emph{\bibinfo{journal}{Proc. Nat. Ac. Sci.}}
  \textbf{\bibinfo{volume}{105}}, \bibinfo{pages}{18681--18686}
  (\bibinfo{year}{2008}).

\bibitem{larson2020conical}
\bibinfo{author}{Larson, J.}, \bibinfo{author}{Sj{\"o}qvist, E.} \&
  \bibinfo{author}{{\"O}hberg, P.}
\newblock \emph{\bibinfo{title}{Conical Intersections in Physics}}
  (\bibinfo{publisher}{Springer}, \bibinfo{year}{2020}).

\bibitem{yarkony1996diabolical}
\bibinfo{author}{Yarkony, D.~R.}
\newblock \bibinfo{title}{Diabolical conical intersections}.
\newblock \emph{\bibinfo{journal}{Rev. Mod. Phys.}}
  \textbf{\bibinfo{volume}{68}}, \bibinfo{pages}{985} (\bibinfo{year}{1996}).

\bibitem{berry1984quantal}
\bibinfo{author}{Berry, M.~V.}
\newblock \bibinfo{title}{Quantal phase factors accompanying adiabatic
  changes}.
\newblock \emph{\bibinfo{journal}{Proc. R. Soc. Lond. A.}}
  \textbf{\bibinfo{volume}{392}}, \bibinfo{pages}{45--57}
  (\bibinfo{year}{1984}).

\bibitem{farag2016probing}
\bibinfo{author}{Farag, M.~H.}, \bibinfo{author}{Jansen, T.~L.} \&
  \bibinfo{author}{Knoester, J.}
\newblock \bibinfo{title}{Probing the interstate coupling near a conical
  intersection by optical spectroscopy}.
\newblock \emph{\bibinfo{journal}{Phys. Chem. Letters}}
  \textbf{\bibinfo{volume}{7}}, \bibinfo{pages}{3328--3334}
  (\bibinfo{year}{2016}).

\bibitem{koppel1983ultrafast}
\bibinfo{author}{K{\"o}ppel, H.}
\newblock \bibinfo{title}{Ultrafast non-radiative decay via conical
  intersections of molecular potential-energy surfaces: C2h4+}.
\newblock \emph{\bibinfo{journal}{Chem. Phys.}} \textbf{\bibinfo{volume}{77}},
  \bibinfo{pages}{359--375} (\bibinfo{year}{1983}).

\bibitem{chen2019mapping}
\bibinfo{author}{Chen, L.}, \bibinfo{author}{Gelin, M.~F.},
  \bibinfo{author}{Zhao, Y.} \& \bibinfo{author}{Domcke, W.}
\newblock \bibinfo{title}{Mapping of wave packet dynamics at conical
  intersections by time-and frequency-resolved fluorescence spectroscopy: A
  computational study}.
\newblock \emph{\bibinfo{journal}{J. Phys. Chem. Lett.}}
  \textbf{\bibinfo{volume}{10}}, \bibinfo{pages}{5873--5880}
  (\bibinfo{year}{2019}).

\bibitem{blatt2012quantum}
\bibinfo{author}{Blatt, R.} \& \bibinfo{author}{Roos, C.~F.}
\newblock \bibinfo{title}{Quantum simulations with trapped ions}.
\newblock \emph{\bibinfo{journal}{Nat. Phys.}} \textbf{\bibinfo{volume}{8}},
  \bibinfo{pages}{277--284} (\bibinfo{year}{2012}).

\bibitem{lloyd1996universal}
\bibinfo{author}{Lloyd, S.}
\newblock \bibinfo{title}{Universal quantum simulators}.
\newblock \emph{\bibinfo{journal}{Science}} \textbf{\bibinfo{volume}{273}},
  \bibinfo{pages}{1073--1078} (\bibinfo{year}{1996}).

\bibitem{macdonell2021analog}
\bibinfo{author}{MacDonell, R.~J.} \emph{et~al.}
\newblock \bibinfo{title}{Analog quantum simulation of chemical dynamics}.
\newblock \emph{\bibinfo{journal}{Chem. Sci.}} \textbf{\bibinfo{volume}{12}},
  \bibinfo{pages}{9794--9805} (\bibinfo{year}{2021}).

\bibitem{gambetta2021exploring}
\bibinfo{author}{Gambetta, F.~M.}, \bibinfo{author}{Zhang, C.},
  \bibinfo{author}{Hennrich, M.}, \bibinfo{author}{Lesanovsky, I.} \&
  \bibinfo{author}{Li, W.}
\newblock \bibinfo{title}{Exploring the many-body dynamics near a conical
  intersection with trapped rydberg ions}.
\newblock \emph{\bibinfo{journal}{Phys. Rev. Lett.}}
  \textbf{\bibinfo{volume}{126}}, \bibinfo{pages}{233404}
  (\bibinfo{year}{2021}).

\bibitem{wuster2011conical}
\bibinfo{author}{W{\"u}ster, S.}, \bibinfo{author}{Eisfeld, A.} \&
  \bibinfo{author}{Rost, J.}
\newblock \bibinfo{title}{Conical intersections in an ultracold gas}.
\newblock \emph{\bibinfo{journal}{Phys. Rev. Lett.}}
  \textbf{\bibinfo{volume}{106}}, \bibinfo{pages}{153002}
  (\bibinfo{year}{2011}).

\bibitem{wuster2018rydberg}
\bibinfo{author}{W{\"u}ster, S.} \& \bibinfo{author}{Rost, J.~M.}
\newblock \bibinfo{title}{Rydberg aggregates}.
\newblock \emph{\bibinfo{journal}{J. Phys. B.}} \textbf{\bibinfo{volume}{51}},
  \bibinfo{pages}{032001} (\bibinfo{year}{2018}).

\bibitem{macdonell2022predicting}
\bibinfo{author}{MacDonell, R.~J.} \emph{et~al.}
\newblock \bibinfo{title}{Predicting molecular vibronic spectra using
  time-domain analog quantum simulation}.
\newblock \emph{\bibinfo{journal}{arXiv preprint arXiv:2209.06558}}
  (\bibinfo{year}{2022}).

\bibitem{omiya2022analytical}
\bibinfo{author}{Omiya, K.} \emph{et~al.}
\newblock \bibinfo{title}{Analytical energy gradient for state-averaged
  orbital-optimized variational quantum eigensolvers and its application to a
  photochemical reaction}.
\newblock \emph{\bibinfo{journal}{J. Chem. Theory Comput.}}
  \textbf{\bibinfo{volume}{18}}, \bibinfo{pages}{741--748}
  (\bibinfo{year}{2022}).

\bibitem{tamiya2021calculating}
\bibinfo{author}{Tamiya, S.}, \bibinfo{author}{Koh, S.} \&
  \bibinfo{author}{Nakagawa, Y.~O.}
\newblock \bibinfo{title}{Calculating nonadiabatic couplings and berry's phase
  by variational quantum eigensolvers}.
\newblock \emph{\bibinfo{journal}{Phys. Rev. Research}}
  \textbf{\bibinfo{volume}{3}}, \bibinfo{pages}{023244} (\bibinfo{year}{2021}).

\bibitem{wang2022observation}
\bibinfo{author}{Wang, C.~S.} \emph{et~al.}
\newblock \bibinfo{title}{Observation of wave-packet branching through an
  engineered conical intersection}.
\newblock \emph{\bibinfo{journal}{arXiv preprint arXiv:2202.02364}}
  (\bibinfo{year}{2022}).

\bibitem{brown2022direct}
\bibinfo{author}{Brown, C.~D.} \emph{et~al.}
\newblock \bibinfo{title}{Direct geometric probe of singularities in band
  structure}.
\newblock \emph{\bibinfo{journal}{Science}} \textbf{\bibinfo{volume}{377}},
  \bibinfo{pages}{1319--1322} (\bibinfo{year}{2022}).

\bibitem{nam2020ground}
\bibinfo{author}{Nam, Y.} \emph{et~al.}
\newblock \bibinfo{title}{Ground-state energy estimation of the water molecule
  on a trapped-ion quantum computer}.
\newblock \emph{\bibinfo{journal}{npj Quantum Inf.}}
  \textbf{\bibinfo{volume}{6}}, \bibinfo{pages}{1--6} (\bibinfo{year}{2020}).

\bibitem{hempel2018quantum}
\bibinfo{author}{Hempel, C.} \emph{et~al.}
\newblock \bibinfo{title}{Quantum chemistry calculations on a trapped-ion
  quantum simulator}.
\newblock \emph{\bibinfo{journal}{Phys. Rev. X}} \textbf{\bibinfo{volume}{8}},
  \bibinfo{pages}{031022} (\bibinfo{year}{2018}).

\bibitem{porras2012quantum}
\bibinfo{author}{Porras, D.}, \bibinfo{author}{Ivanov, P.~A.} \&
  \bibinfo{author}{Schmidt-Kaler, F.}
\newblock \bibinfo{title}{Quantum simulation of the cooperative jahn-teller
  transition in 1d ion crystals}.
\newblock \emph{\bibinfo{journal}{Phys. Rev. Lett.}}
  \textbf{\bibinfo{volume}{108}}, \bibinfo{pages}{235701}
  (\bibinfo{year}{2012}).

\bibitem{gorman2018engineering}
\bibinfo{author}{Gorman, D.~J.} \emph{et~al.}
\newblock \bibinfo{title}{Engineering vibrationally assisted energy transfer in
  a trapped-ion quantum simulator}.
\newblock \emph{\bibinfo{journal}{Phys. Rev. X}} \textbf{\bibinfo{volume}{8}},
  \bibinfo{pages}{011038} (\bibinfo{year}{2018}).

\bibitem{richerme2022quantum}
\bibinfo{author}{Richerme, P.} \emph{et~al.}
\newblock \bibinfo{title}{Quantum computation of hydrogen bond dynamics and
  vibrational spectra}.
\newblock \emph{\bibinfo{journal}{arXiv preprint arXiv:2204.08571}}
  (\bibinfo{year}{2022}).

\bibitem{monroe2021programmable}
\bibinfo{author}{Monroe, C.} \emph{et~al.}
\newblock \bibinfo{title}{Programmable quantum simulations of spin systems with
  trapped ions}.
\newblock \emph{\bibinfo{journal}{Rev. Mod. Phys.}}
  \textbf{\bibinfo{volume}{93}}, \bibinfo{pages}{025001}
  (\bibinfo{year}{2021}).

\bibitem{nguyen2022digital}
\bibinfo{author}{Nguyen, N.~H.} \emph{et~al.}
\newblock \bibinfo{title}{Digital quantum simulation of the schwinger model and
  symmetry protection with trapped ions}.
\newblock \emph{\bibinfo{journal}{PRX {Q}uantum}} \textbf{\bibinfo{volume}{3}},
  \bibinfo{pages}{020324} (\bibinfo{year}{2022}).

\bibitem{georgescu2014quantum}
\bibinfo{author}{Georgescu, I.~M.}, \bibinfo{author}{Ashhab, S.} \&
  \bibinfo{author}{Nori, F.}
\newblock \bibinfo{title}{Quantum simulation}.
\newblock \emph{\bibinfo{journal}{Rev. Mod. Phys.}}
  \textbf{\bibinfo{volume}{86}}, \bibinfo{pages}{153} (\bibinfo{year}{2014}).

\bibitem{johansson2012qutip}
\bibinfo{author}{Johansson, J.~R.}, \bibinfo{author}{Nation, P.~D.} \&
  \bibinfo{author}{Nori, F.}
\newblock \bibinfo{title}{Qutip: An open-source python framework for the
  dynamics of open quantum systems}.
\newblock \emph{\bibinfo{journal}{Comp. Phys. Comm.}}
  \textbf{\bibinfo{volume}{183}}, \bibinfo{pages}{1760--1772}
  (\bibinfo{year}{2012}).

\bibitem{berry1990anticipations}
\bibinfo{author}{Berry, M.} \emph{et~al.}
\newblock \bibinfo{title}{Anticipations of the geometric phase}.
\newblock \emph{\bibinfo{journal}{Physics Today}}
  \textbf{\bibinfo{volume}{43}}, \bibinfo{pages}{34--40}
  (\bibinfo{year}{1990}).

\bibitem{longuet1958studies}
\bibinfo{author}{Longuet-Higgins, H.~C.}, \bibinfo{author}{{\"O}pik, U.},
  \bibinfo{author}{Pryce, M. H.~L.} \& \bibinfo{author}{Sack, R.}
\newblock \bibinfo{title}{Studies of the jahn-teller effect. ii. the dynamical
  problem}.
\newblock \emph{\bibinfo{journal}{Proc. R. Soc. Lond. A.}}
  \textbf{\bibinfo{volume}{244}}, \bibinfo{pages}{1--16}
  (\bibinfo{year}{1958}).

\bibitem{manchon2015new}
\bibinfo{author}{Manchon, A.}, \bibinfo{author}{Koo, H.~C.},
  \bibinfo{author}{Nitta, J.}, \bibinfo{author}{Frolov, S.} \&
  \bibinfo{author}{Duine, R.}
\newblock \bibinfo{title}{New perspectives for rashba spin--orbit coupling}.
\newblock \emph{\bibinfo{journal}{Nat. Mater.}} \textbf{\bibinfo{volume}{14}},
  \bibinfo{pages}{871--882} (\bibinfo{year}{2015}).

\bibitem{Lin_2011}
\bibinfo{author}{Lin, Y.-J.}, \bibinfo{author}{Jim{\'{e}}nez-Garc{\'{\i}}a, K.}
  \& \bibinfo{author}{Spielman, I.~B.}
\newblock \bibinfo{title}{Spin{\textendash}orbit-coupled
  bose{\textendash}einstein condensates}.
\newblock \emph{\bibinfo{journal}{Nature}} \textbf{\bibinfo{volume}{471}},
  \bibinfo{pages}{83--86} (\bibinfo{year}{2011}).
\newblock \urlprefix\url{https://doi.org/10.1038%2Fnature09887}.

\bibitem{wang2020high}
\bibinfo{author}{Wang, Y.} \emph{et~al.}
\newblock \bibinfo{title}{High-fidelity two-qubit gates using a
  microelectromechanical-system-based beam steering system for individual qubit
  addressing}.
\newblock \emph{\bibinfo{journal}{Phys. Rev. Lett.}}
  \textbf{\bibinfo{volume}{125}}, \bibinfo{pages}{150505}
  (\bibinfo{year}{2020}).

\bibitem{jia2022determination}
\bibinfo{author}{Jia, Z.} \emph{et~al.}
\newblock \bibinfo{title}{Determination of multi-mode motional quantum states
  in a trapped ion system}.
\newblock \emph{\bibinfo{journal}{arXiv preprint arXiv:2205.11444}}
  (\bibinfo{year}{2022}).

\bibitem{gerritsma2010quantum}
\bibinfo{author}{Gerritsma, R.} \emph{et~al.}
\newblock \bibinfo{title}{Quantum simulation of the dirac equation}.
\newblock \emph{\bibinfo{journal}{Nature}} \textbf{\bibinfo{volume}{463}},
  \bibinfo{pages}{68--71} (\bibinfo{year}{2010}).

\bibitem{katz2022programmable}
\bibinfo{author}{Katz, O.} \& \bibinfo{author}{Monroe, C.}
\newblock \bibinfo{title}{Programmable quantum simulations of bosonic systems
  with trapped ions}.
\newblock \emph{\bibinfo{journal}{arXiv preprint arXiv:2207.13653}}
  (\bibinfo{year}{2022}).

\bibitem{katz2022programmable2}
\bibinfo{author}{Katz, O.}, \bibinfo{author}{Cetina, M.} \&
  \bibinfo{author}{Monroe, C.}
\newblock \bibinfo{title}{Programmable n-body interactions with trapped ions}.
\newblock \emph{\bibinfo{journal}{arXiv preprint arXiv:2207.10550}}
  (\bibinfo{year}{2022}).

\bibitem{lemmer2018trapped}
\bibinfo{author}{Lemmer, A.} \emph{et~al.}
\newblock \bibinfo{title}{A trapped-ion simulator for spin-boson models with
  structured environments}.
\newblock \emph{\bibinfo{journal}{New J. Phys.}} \textbf{\bibinfo{volume}{20}},
  \bibinfo{pages}{073002} (\bibinfo{year}{2018}).

\bibitem{roos2008ion}
\bibinfo{author}{Roos, C.~F.}
\newblock \bibinfo{title}{Ion trap quantum gates with amplitude-modulated laser
  beams}.
\newblock \emph{\bibinfo{journal}{New J. Phys.}} \textbf{\bibinfo{volume}{10}},
  \bibinfo{pages}{013002} (\bibinfo{year}{2008}).

\bibitem{leung2018robust}
\bibinfo{author}{Leung, P.~H.} \emph{et~al.}
\newblock \bibinfo{title}{Robust 2-qubit gates in a linear ion crystal using a
  frequency-modulated driving force}.
\newblock \emph{\bibinfo{journal}{Phys. Rev. Lett.}}
  \textbf{\bibinfo{volume}{120}}, \bibinfo{pages}{020501}
  (\bibinfo{year}{2018}).

\bibitem{batista2001generalized}
\bibinfo{author}{Batista, C.} \& \bibinfo{author}{Ortiz, G.}
\newblock \bibinfo{title}{Generalized jordan-wigner transformations}.
\newblock \emph{\bibinfo{journal}{Phys. Rev. Lett.}}
  \textbf{\bibinfo{volume}{86}}, \bibinfo{pages}{1082} (\bibinfo{year}{2001}).

\bibitem{seeley2012bravyi}
\bibinfo{author}{Seeley, J.~T.}, \bibinfo{author}{Richard, M.~J.} \&
  \bibinfo{author}{Love, P.~J.}
\newblock \bibinfo{title}{The bravyi-kitaev transformation for quantum
  computation of electronic structure}.
\newblock \emph{\bibinfo{journal}{J. Chem. Phys.}}
  \textbf{\bibinfo{volume}{137}}, \bibinfo{pages}{224109}
  (\bibinfo{year}{2012}).

\bibitem{valahu2022direct}
\bibinfo{author}{Valahu, C.~H.} \emph{et~al.}
\newblock \bibinfo{title}{Direct observation of geometric phase in dynamics
  around a conical intersection}.
\newblock \emph{\bibinfo{journal}{arXiv preprint arXiv:2211.07320}}
  (\bibinfo{year}{2022}).

\bibitem{olmschenk2007manipulation}
\bibinfo{author}{Olmschenk, S.} \emph{et~al.}
\newblock \bibinfo{title}{Manipulation and detection of a trapped yb+ hyperfine
  qubit}.
\newblock \emph{\bibinfo{journal}{Phys. Rev. A}} \textbf{\bibinfo{volume}{76}},
  \bibinfo{pages}{052314} (\bibinfo{year}{2007}).

\bibitem{revelle2020phoenix}
\bibinfo{author}{Revelle, M.~C.}
\newblock \bibinfo{title}{Phoenix and peregrine ion traps}.
\newblock \emph{\bibinfo{journal}{arXiv preprint arXiv:2009.02398}}
  (\bibinfo{year}{2020}).

\bibitem{debnath2016programmable}
\bibinfo{author}{Debnath, S.}
\newblock \emph{\bibinfo{title}{A programmable five qubit quantum computer
  using trapped atomic ions}}.
\newblock Ph.D. thesis, \bibinfo{school}{University of Maryland, College Park}
  (\bibinfo{year}{2016}).

\bibitem{hayes2010entanglement}
\bibinfo{author}{Hayes, D.} \emph{et~al.}
\newblock \bibinfo{title}{Entanglement of atomic qubits using an optical
  frequency comb}.
\newblock \emph{\bibinfo{journal}{Phys. Rev. Lett.}}
  \textbf{\bibinfo{volume}{104}}, \bibinfo{pages}{140501}
  (\bibinfo{year}{2010}).

\bibitem{wineland1998experimental2}
\bibinfo{author}{Wineland, D.~J.} \emph{et~al.}
\newblock \bibinfo{title}{Experimental issues in coherent quantum-state
  manipulation of trapped atomic ions}.
\newblock \emph{\bibinfo{journal}{Journal of research of the National Institute
  of Standards and Technology}} \textbf{\bibinfo{volume}{103}},
  \bibinfo{pages}{259} (\bibinfo{year}{1998}).

\bibitem{Leibfried:2003}
\bibinfo{author}{Leibfried, D.}, \bibinfo{author}{Blatt, R.},
  \bibinfo{author}{Monroe, C.} \& \bibinfo{author}{Wineland, D.}
\newblock \bibinfo{title}{Quantum dynamics of single trapped ions}.
\newblock \emph{\bibinfo{journal}{Rev. Mod. Phys.}}
  \textbf{\bibinfo{volume}{75}}, \bibinfo{pages}{281--324}
  (\bibinfo{year}{2003}).
\newblock \urlprefix\url{https://link.aps.org/doi/10.1103/RevModPhys.75.281}.

\bibitem{wineland1998experimental}
\bibinfo{author}{Wineland, D.~J.} \emph{et~al.}
\newblock \bibinfo{title}{Experimental primer on the trapped ion quantum
  computer}.
\newblock \emph{\bibinfo{journal}{Prog. Phys.}} \textbf{\bibinfo{volume}{46}},
  \bibinfo{pages}{363--390} (\bibinfo{year}{1998}).

\end{thebibliography}

\clearpage

\section{Supplementary Material}

\subsection{Mapping to the Trapped Ion Hamiltonian}

\par
Much of this follows the work of \cite{Leibfried:2003}. We start with one ion, but the extension to multiple ions is fairly obvious. Ignoring the unused axial mode (which has a much lower frequency and therefore is not coupled into), the one ion Hamiltonian can be written as follows:
\begin{equation}
    \hat{H}_{ion} = \frac{ \omega_0}{2} \hat{\sigma}_z +  \nu_x \hat{n}_x +  \nu_y \hat{n}_y + \hat{H}^{(i)},
\end{equation}
where $\omega_0$ is the energy difference between chosen qubit states, and $\nu_x$ and $\nu_y$ are the vibrational frequencies of the $x$ and $y$ modes, respectively. Here, the motion of the ion has been approximated as a harmonic oscillator, which is a reasonable assumption if the micromotion caused by the RF drive frequency can be ignored. The interaction between the motional modes and the electronic states of the ion, mediated via four different lasers, is:
\begin{equation}
    \hat{H}^{(i)} = \sum_{j=1,2,  q = x,y} \frac{ \Omega_{j, q}}{2}\sigma_x \big(e^{i(k_{j, q} \hat{q} - \omega_{j,q} t + \phi_{j, q})} + h.c. \big),
\end{equation}
where $\Omega_{j, q}$, $k_{j, q}$, $\omega_{j, q}$, $\phi_{j, q}$ are the electric field strength, wave vector, frequency, and phase of the $j$th beam intended to couple into mode $q$. Here, two of the lasers interact with the $x$ motion of the ion and the other two interact with the $y$ motion. Moving into the interaction picture, and performing the rotating wave approximation, we get the following Hamiltonian:
\begin{widetext}
\begin{equation}
    \begin{aligned}
        \hat{H}_{int} & = \exp\big(-i (\hat{H}_{ion} - \hat{H}^{(i)})t\big) \hat{H}^{(i)} \exp\big(i (\hat{H}_{ion} - \hat{H}^{(i)})t\big) \\
        & \approx \sum_{j=1,2,  q = x,y} \big(\frac{ \Omega_{j, q}}{2}\big)\bigg(\hat{\sigma}_{+}\exp\big(i[\eta_{j, q}(\hat{a}_q e^{-i \nu_q t} + \hat{a}_q^{\dagger} e^{i \nu_q t}) - \delta_{j, q} t + \phi_{j, q}]\big) + h.c.\bigg),
    \end{aligned}
\end{equation}
\end{widetext}
where $\delta_{j, q}$ is the detuning between the laser and qubit resonance, and $\eta_{j, q}$ is the Lamb-Dicke parameter taking into account the motional coupling between the laser and the ion. In the Lamb-Dicke regime, i.e. $\eta\braket{2n + 1} \ll 1$, we can approximate the exponential operators to the first order, leaving us with the following approximate interaction Hamiltonian:
\begin{widetext}
\begin{equation}
    \hat{H}_{int} \approx \sum_{j=1,2,  q = x,y} \big(\frac{ \Omega_{j, q}}{2}\big)\bigg(\hat{\sigma}_{+}(1 + i\eta_{j, q}(\hat{a}_q e^{-i \nu_q t} + \hat{a}_q^{\dagger} e^{i \nu_q t})) \exp\big(i[-\delta_{j, q} t + \phi_{j, q}]\big) + h.c.\bigg).
\end{equation}
\end{widetext}
If we properly set the phase and detuning of each of our lasers, and perform a second rotating wave approximation, we can obtain the following Hamiltonian:
\begin{equation}
    \hat{H}_{int} \approx \big(\frac{ \eta \Omega}{2}\big)\bigg(\hat{\sigma}_x (\hat{a}_x e^{-i \nu t} + \hat{a}_x^{\dagger} e^{i \nu t}) + \hat{\sigma}_y (\hat{a}_y e^{-i \nu t} + \hat{a}_y^{\dagger} e^{i \nu t})\bigg).
\end{equation}
This is clearly very close to the desired Hamiltonian in Eq. \ref{equation:quantum_JT}.  All that remains is to rotate slightly out of the full interaction picture with a number operators in both modes proportional to $\nu$.  The result is the following effective Hamiltonian:
\begin{equation}
    \hat{H}_{eff} \approx  \nu (\hat{n}_x + \hat{n}_y) + \big(\frac{ \eta \Omega}{2}\big)\bigg(\hat{\sigma}_x (\hat{a}_x + \hat{a}_x^{\dagger}) + \hat{\sigma}_y (\hat{a}_y + \hat{a}_y^{\dagger})\bigg).
\end{equation}
This is exactly what we want.  Unfortunately we have made several approximations to obtain this result. To account for these during classical simulation of our experiment, we used the following less approximate Hamiltonian instead:
\begin{widetext}
\begin{equation}
    \begin{aligned}
        \hat{H}_{noisy} \approx & \frac{ \Delta_z}{2}\hat{\sigma}_z +  \nu (\hat{n}_x + \hat{n}_y) + \\ & \big(\frac{ \eta \Omega}{2}\big)\bigg(\hat{\sigma}_x (\hat{a}_x + \hat{a}_x^{\dagger}) + \hat{\sigma}_y (\hat{a}_y + \hat{a}_y^{\dagger}) + \hat{\sigma}_x ((\hat{a}_y + \hat{a}_y^{\dagger})\cos(\Delta_{xy} t) + \\
        & i(\hat{a}_y - \hat{a}_y^{\dagger})\sin(\Delta_{xy} t)) + \hat{\sigma}_y ((\hat{a}_x + \hat{a}_x^{\dagger})\cos(\Delta_{xy} t) - i(\hat{a}_x - \hat{a}_x^{\dagger})\sin(\Delta_{xy} t)) \bigg).
    \end{aligned}
\end{equation}
\end{widetext}
This takes into account improper detuning in our lasers via $\Delta_z$, as well as any cross coupling between lasers meant for the x mode with the y mode, and vice versa, where $\Delta_{xy}$ is the frequency difference between the modes.  This noise was approximated away via the second rotating wave approximation we made.  Lastly, we took into account heating and phase noise caused by micromotion and other sources by plugging this Hamiltonian into the full Lindblad master equation, $\dot{\rho} = \frac{1}{i } \big[\hat{H}_{noisy}, \rho\big] + \frac{1}{2}\sum_{j} \gamma_j (2 L_j \rho L_j^{\dagger} - L_j^{\dagger} L_j \rho - \rho L_j^{\dagger} L_j)$, where $L_j \in \{\hat{a}_x, \hat{a}_x^{\dagger}, \hat{n}_x, \hat{a}_y, \hat{a}_y^{\dagger}, \hat{n}_y\}$.

\subsection{Effect of Off-Resonant Coupling on Measurement}

In our system, we used one of the nearest neighbors of the center ion to measure the normal mode distributions of the ions.  This ion couples to all modes, unlike the center ion, and we need to include off-resonant effects on the "Fourier-Push".  We start by including one extra mode, using the following effective Hamiltonian:

\begin{equation}
    H = \frac{\eta_1 \Omega}{2}\sigma_x (a_1 + a_1^{\dagger}) + \frac{\eta_2 \Omega}{2}\sigma_x (a_2 e^{-i \Delta t} + a_2^{\dagger}e^{i \Delta t}), 
\end{equation}
where $\eta_1$ and $\eta_2$ are the Lamb-Dicke parameters for the two respective modes, $\Omega$ is the carrier Rabi frequency, $\Delta$ is the detuning from mode $2$, and we are trying measure mode $1$. We note from \cite{roos2008ion} that for any Hamiltonian that can be written as $i(\gamma(t) a + \gamma^*(t) a^{\dagger})\mathcal{O}$, the unitary evolution is $U(t) = \hat{D}(\alpha(t)\mathcal{O})exp(i\Phi(t) \mathcal{O}^2)$, where $\hat{D}$ is the displacement operator, $\alpha(t) = \int_0^t dt' \gamma(t')$, $\Phi(t) = Im[\int_0^t dt' \gamma(t') \int_0^{t'}dt''\gamma(t'')]$, and $\mathcal{O}$ is any arbitrary operator. In our case, $\mathcal{O} = \sigma_x$, and therefore $\mathcal{O}^2$ is the identity, so the second term is simply a global phase we can ignore.  The two parts of the Hamiltonian that couple into different modes commute, so we can write the unitary evolution as:

\begin{widetext}
\begin{equation}
    U(t) = \exp\big(\frac{-i \eta_1 \Omega t}{2}\sigma_x (a_1 + a_1^{\dagger})\big) \exp\big(-i \frac{\eta_2 \Omega}{\Delta} \sin(\Delta t/2) \sigma_x (a_2 e^{-i\Delta/2} + a_2^{\dagger}e^{i\Delta/2})\big).
\end{equation}
\end{widetext}
By expanding in a similar way, we reach the following expectation for our $\sigma_z$ projection measurement:
\begin{widetext}
\begin{equation}
    \begin{aligned}
        & \langle U(t)^{\dagger} \sigma_z U(t) \rangle = \\ & \langle \sigma_z \big(\cos(\sqrt{2}\eta_1 \Omega t x_1) \cos(2\sqrt{2}\frac{\eta_2 \Omega}{\Delta} \sin(\Delta t/2) (x_2 \cos(\Delta t/2) + p_2 \sin(\Delta t / 2))) \\ & + \sin(\sqrt{2}\eta_1 \Omega t x_1) \sin(2\sqrt{2}\frac{\eta_2 \Omega}{\Delta} \sin(\Delta t/2) (x_2 \cos(\Delta t/2) + p_2 \sin(\Delta t / 2)))\big) \rangle + \\ & \langle \sigma_y \big(\cos(\sqrt{2}\eta_1 \Omega t x_1) \sin(2\sqrt{2}\frac{\eta_2 \Omega}{\Delta} \sin(\Delta t/2) (x_2 \cos(\Delta t/2) + p_2 \sin(\Delta t / 2))) \\ & + \sin(\sqrt{2}\eta_1 \Omega t x_1) \cos(2\sqrt{2}\frac{\eta_2 \Omega}{\Delta} \sin(\Delta t/2) (x_2 \cos(\Delta t/2) + p_2 \sin(\Delta t / 2)))\big) \rangle.
    \end{aligned}
\end{equation}
\end{widetext}
We see that we are essentially taking a secondary Fourier transform at this point, excecpt with an oscillating k-number.  This has the effect of modulating the original singal. At this point, we make the approximation that the off-resonant mode is in a thermal state.  This helps us in two ways, first by removing any odd portions of new modulating terms, and second by allowing us to ignore any difference between $x_2$ and $p_2$.  The new expectation value is:
\begin{widetext}
\begin{equation}
    \begin{aligned}
        & \langle U(t)^{\dagger} \sigma_z U(t) \rangle = \langle \cos(2\sqrt{2}\frac{\eta_2 \Omega}{\Delta} \sin(\Delta t/2) x_2) \big( \sigma_z \cos(\sqrt{2}\eta_1 \Omega t x_1) + \sigma_y \sin(\sqrt{2}\eta_1 \Omega t x_1)\big) \rangle \\ & = \int dx_1 dx_2 |\psi_1(x_1)|^2 |\psi_2(x_2)|^2 \cos(2\sqrt{2}\frac{\eta_2 \Omega}{\Delta} \sin(\Delta t/2) x_2) \big( \sigma_z \cos(\sqrt{2}\eta_1 \Omega t x_1) + \sigma_y \sin(\sqrt{2}\eta_1 \Omega t x_1)\big).
    \end{aligned}
\end{equation}
\end{widetext}

Because it's in a thermal state, we assume that $|\psi_2(x_2)|^2 \propto \exp(-x_2^2/2 \sigma^2)$ where $\sigma$ depends on how well cooled the mode is.  By performing the Fourier transform, we get:
\begin{widetext}
\begin{equation}
    \begin{aligned}
        & \langle U(t)^{\dagger} \sigma_z U(t) \rangle = \\  & \int dx_1 |\psi_1(x_1)|^2 \exp(-4 \sigma^2 \frac{\eta_2^2 \Omega^2}{\Delta^2} \sin^2(\frac{\Delta t}{2})) \big( \braket{\sigma_z} \cos(\sqrt{2}\eta_1 \Omega t x_1) + \braket{\sigma_y} \sin(\sqrt{2}\eta_1 \Omega t x_1)\big).
    \end{aligned}
\end{equation}
\end{widetext}
We see that the signal is modulated by a Gaussian-like wave-packet, except the argument is a sine function. The effect of this is to add Gaussian noise to the inverse Fourier transform (which in turn appears to be a large thermal background).  In theory, the signal could be demodulated if the exact parameters of $|\psi_2(x_2)|^2$ are known, but this is difficult enough for one extra mode and in one dimension.  In our experiment, the measurement in the effective x-direction is close enough to one mode to be affected, while the measurement of the y-direction is close enough to two modes.  The final signal would have to be demodulated in three different ways.

\subsection{Sensitivity to Noise}

While the one ion case is intuitive to map to for this experiment, it is not as resistant to noise as a multi ion simulation \cite{wineland1998experimental, wineland1998experimental2}.  Heating rates in the center of mass modes of ion chains tend to be higher than those of other modes ($>$ 100 quanta/sec in our set up), and the only mode available to a single ion simulation is the center of mass mode.  With two ions, we have the tilt mode available in the x and y axes, which corresponds to the ions oscillating exactly out of phase with each other and can have a heating rate as low as an order of magnitude smaller than that of the center of mass mode.  

With a five ion setup, we have even less sensitive modes to utilize \cite{debnath2016programmable}.  The best is the zigzag mode, which corresponds to nearest neighbor ions moving out of phase with each other.  Due to our choice of ions to use (the middle ion and the one directly to the left), the next best available mode is the so-called third mode, which corresponds to the middle three ions moving out of phase with the edge ions.  Both of these modes have measured heating rates of less than one per second, and motional dephasing times on the order of 8 ms, much longer than our adiabatic evolution.

\subsection{Justification of Different Evolution}

\begin{figure*}[ht!]
    \begin{tabular}{cc}
         \includegraphics[scale=0.3]{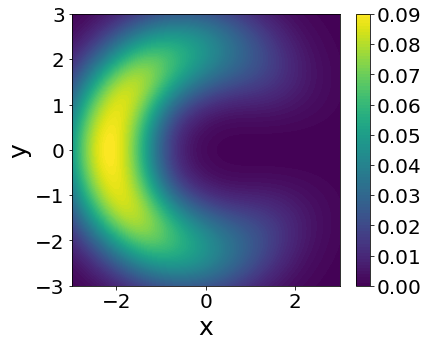} 
         (a) 
         \includegraphics[scale=0.3]{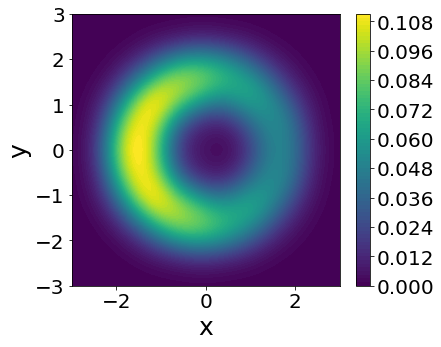} 
         (b) 
         \includegraphics[scale=0.3]{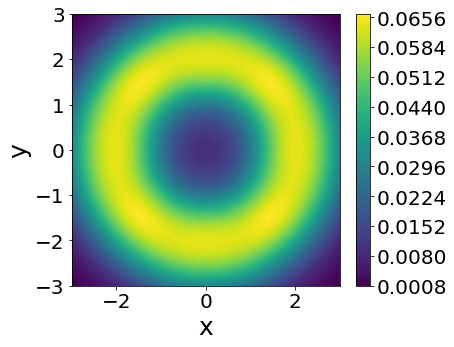} 
         (c) 
    \end{tabular}
    \caption{\textbf{a}, Simulation results of adiabatic evolution when CI is present, where $\Omega = 2 \pi \times 10$ kHz and $\nu = 2 \pi \times 3$ kHz for both modes. \textbf{b}, Same simulation results, but switching to the non-CI Hamiltonian midway through the evolution.  The geometric phase interference is less prominent.  \textbf{c}, Using the non-CI Hamiltonian the whole way through.  There is no sign of geometric phase interference. Slight difference in outermost points most likely stem from non-adiabatic behavior when switching Hamiltonians. }
    \label{fig:removing_geo_phase}
\end{figure*}

Our evolution is not the same as the natural choice of a circle around the intersection.  We turn on both the x-mode interaction and the y-mode interaction at the same time.  This immediately brings in the argument that by starting at the degeneracy point, or the point where the notion of adiabaticity makes no sense, we invalidate one of the key requirements of geometric phase.  Our intuitive argument that this still produces geometric phase interference is that by starting in the $\ket{+}$ state, we have broken a symmetry.  This pushes the ion to the negative direction in the x-mode in the beginning.  After that, the conical intersection is still at the center of the potential energy surface, and the whole wavefunction is still in the ground state of the Hamiltonian.  Thus, from this point onward, all the normal arguments apply, and the only thing stopping the ion from spreading to the other side over time has to be a geometric phase, because the dynamic phase on the wavefunction is the same for all positions.  

More rigorously, we note the following equations of motion given by the dynamics of the Hamiltonian:

\begin{equation}
    \begin{aligned}
        \braket{\ddot{x}} = -\nu^2 \braket{x} - \frac{\Omega \nu}{\sqrt{2}}\braket{\sigma_x}, \\
        \braket{\ddot{y}} = -\nu^2 \braket{y} - \frac{\Omega \nu}{\sqrt{2}}\braket{\sigma_y}.
    \end{aligned}
\end{equation}
For very small times, $\braket{\sigma_x} \approx 1$ and thus $\nu |\braket{x}| \ll \frac{\Omega}{2}|\braket{\sigma_x}|$, meaning there's an initial push in the negative direction of the whole wavefunction of about $\braket{x} \approx -\frac{\nu \Omega}{2\sqrt{2}}t^2$. By mapping $x + iy = z$, we can perform a contour integral and utilize the residue theorem. This is particularly convenient because Eq. \ref{equation:xy_form} takes the form:

\begin{equation}
    \gamma_{\pm} = \frac{1}{4i}\int^{z_f}_{z_0}  \big(\frac{dz}{z} - \frac{dz^*}{z^*}\big).
\end{equation}
This integral famously evaluates to $\pm \pi$ if it makes a loop around the singularity at the center point $z = 0$, depending on the direction of travel, and evaluates to $\pm \frac{\pi}{2}$ if it travels exactly halfway around the singularity. Based on our initial push in the negative direction, our parameterization can roughly be written as $x = \cos(\theta) + \epsilon$ and $y = \sin(\theta)$ for $\theta_0 = \pm \pi$ and $\theta_f = 0$, where $\epsilon \approx -\frac{\nu \Omega}{2\sqrt{2}}t^2$. Thus, by encompassing the intersection even just barely, we have achieved a non-trivial geometric phase. 

We also noted that there are symmetries maintained throughout the evolution by choosing this path.  In particular, the quantity $L_z + \frac{1}{2}\sigma_z$ is conserved throughout the evolution, something that cannot be said about the naive adiabatic evolution.  This means that the evolution cannot make transitions between eigensupaces that do not share this value.  We believe this helps maintain adiabatcity despite a much shorter evolution time.  Based on classical simulations, we found that if we take 25 ms as an appropriate amount of time to evolve the state using the naive evolution method, and we compare that to the same evolution method but with only 330 $\mu$s to evolve, the fidelity between the two is about $80\%$.  If we instead use the symmetric method explained here, the same evolution time of 330 $\mu$s give us a fidelity of over $96 \%$.

For completeness, we compared three different evolutions via classical simulation.  The first is the planned evolution with the ideal Hamiltonian and the conical intersection present, and the final state is shown in Fig.\ref{fig:removing_geo_phase}a. In the evolution represented by Fig.\ref{fig:removing_geo_phase}b, halfway through we switched to the Hamiltonian $H(t) = \nu n_x + \nu n_y - \frac{\Omega(t)}{2}\sqrt{(a_x + a_x^{\dagger})^2 + (a_y + a_y^{\dagger})^2}$, which does not have a conical intersection, but has the same adiabatic potential energy surface as the lower half of the ideal Hamiltonian.  As can be seen, the geometric phase is less present.  Finally, in Fig.\ref{fig:removing_geo_phase}c we used the non-CI Hamiltonian the whole evolution, and there is no sign of geometric phase interference.

\end{document}